\documentclass[aps,prl,twocolumn,groupedaddress]{revtex4}
\usepackage{graphicx}
\usepackage{amsmath}
\usepackage{relsize}
\usepackage{hyperref}
\usepackage{multirow}
\usepackage{color}
\usepackage{hhline}
\usepackage{float}
\restylefloat{table}

\begin{document}

\title{Locating the pseudogap closing point in cuprate superconductors: absence of entrant or reentrant behavior}

\author{Jeffery L. Tallon$^{1,\dag}$, James G. Storey$^1$, John R. Cooper$^2$ and John W. Loram$^{2,\ddag}$}

\affiliation{$^1$Robinson Research Institute, Victoria University of Wellington,
P.O. Box 33436, Lower Hutt 5046, New Zealand.}

\affiliation{$^2$Cavendish Laboratory, Cambridge University, Cambridge CB3 0HE, England.}

\date{\today}
\begin{abstract}
Many current descriptions of the pseudogap in underdoped cuprates envision a doping-dependent transition line $T^*(p)$ which descends monotonically towards zero just beyond optimal doping. There is much debate as to the location of the terminal point $p^*$ where $T^*(p)$ vanishes, whether or not there is a phase transition at  $T^*$ and exactly how $T^*(p)$ behaves below $T_c$ within the superconducting dome. One perspective sees $T^*(p)$ cutting the dome and continuing to descend monotonically to zero at $p^* \approx 0.19$ holes/Cu $-$ referred to here as `entrant behavior'. Another perspective derived from photoemission studies is that $T^*(p)$ intersects the dome near $p \approx 0.23$ holes/Cu then turns back below $T_c$, falling to zero again around $p^* \approx 0.19$ $-$ referred to here as `reentrant behavior'. By examining field-dependent thermodynamic data for Bi$_2$Sr$_2$CaCu$_2$O$_{8+\delta}$ we show that neither entrant nor reentrant behavior is supported. Rather, $p^*$ sharply delimits the pseudogap regime: for $p < 0.19$ the pseudogap is always present, independent of $T$. Similar results are found for Y$_{0.8}$Ca$_{0.2}$Ba$_2$Cu$_3$O$_{7-\delta}$. For both materials $T^*(p)$ is not a temperature but a crossover scale, $\approx E^*(p)/2k_B$, reflecting instead the underlying pseudogap energy $E^*(p)$ which vanishes as $p \rightarrow p^* \approx 0.19$.
\end{abstract}

\pacs{74.25.Bt, 74.40.kb, 74.72.-h}

\maketitle

\section*{1. Introduction}

Hole-doped cuprate superconductors, at and below optimal doping, are characterised by the opening of a partial gap in the electronic density of states, the so-called pseudogap \cite{Norman1,Timusk}, which profoundly affects all spectroscopic properties and, below the transition temperature $T_c$, results in an abrupt crossover from `strong' to `weak' superconductivity \cite{Loram,Bernhard,Grissonnanche}. Recently, it has become evident that the underlying behavior involves a change of the Fermi surface (FS) from large, with area $1+p$, to either Fermi arcs \cite{Norman2} or small hole pockets on the zone diagonal near the antiferromagnetic zone boundary and having area $p$ \cite{Storey1,Storey2,Davis,Hudson,Badoux}. Evidence for this change can be found in angle-resolved photoelectron spectroscopy (ARPES) \cite{Johnson,Damascelli}, quasiparticle interference in scanning tunneling spectroscopy \cite{Davis,Hudson} and the reported crossover observed in the normal state at very high magnetic field from Hall number $n_H = 1+p$ to $n_H = p$ \cite{Badoux,Storey1}. (However, recent high-field Hall measurements on Tl$_2$Ba$_2$CuO$_6$ and Bi$_2$Sr$_2$CuO$_6$ do not support this interpretation \cite{Carrington}) - see Appendix C.

It has long been known that the {\it apparent} characteristic pseudogap temperature, $T^*$, below which pseudogap effects are often reported, falls with increasing hole concentration, $p$, and vanishes at a critical doping, $p^* \approx 0.19$ holes/Cu \cite{Tstar,Naqib}. Despite an intensive search no specific heat anomaly has been reported at $T^*$ \cite{Loram,Cooper}, thus implying that $T^*(p)$ is not a thermodynamic phase transition line. Indeed this line was originally reported as an {\it energy scale}, $E^*(p)$, which descends to zero at $p^*$, not a {\it temperature scale} \cite{Loram1,Loram3,Tallon1,Williams}. Later reports however suggest that there is indeed some kind of mean-field transition occurring in the vicinity of $T^*$ with various order-parameter-like properties observed to vanish there. The measurements include polarised neutron scattering \cite{Bourges}, polar Kerr effect \cite{Xia}, ARPES \cite{Hashimoto}, time-resolved reflectivity \cite{He1}, resonant ultrasound spectroscopy \cite{Shekhter} and susceptibility nematicity \cite{Sato}. The last two techniques were presented as clear evidence of a thermodynamic transition \cite{Shekhter,Sato}, although in the former case we have questioned this interpretation \cite{Cooper} and any thermodynamic effects are evidently extremely weak. In contrast, the pseudogap effects reported from the specific heat are very strong, involving a large suppression of entropy and a consequent suppression of the superconducting jump at $T_{\textrm{c}}$, the condensation energy and the superfluid density \cite{Loram,Bernhard}. This necessarily raises the question as to whether these obviously disparate results are at all related.  Also, of special relevance to the present work, several prominent studies on Bi$_2$Sr$_2$CaCu$_2$O$_{8+\delta}$ (Bi2212) identified the termination point of the pseudogap as located at a much higher doping of $p^* \approx 0.23$, at or near a proposed Lifshitz transition from a hole-like to electron-like Fermi surface \cite{Kaminski}. These include ARPES \cite{Vishik,He2}, Raman\cite{Sacuto1} and transport \cite{Taillefer} studies.

\subsection*{1.1 The problem stated} It is obvious from the above that there is a clear contradiction between these more recent studies and the earlier thermodynamic, spectroscopic and transport studies. In the present work we seek to address and resolve this contradiction.

Additionally, we examine the thermodynamic behaviour in the neighbourhood of $p^*$ within the superconducting dome to search for what we call `entrant' or `reentrant' behavior. If $T^*$ delineates the opening of the pseudogap, as claimed, and if the pseudogap is responsible for a large loss of spectral weight as is obvious from thermodynamic, NMR \cite{Dupree,Alloul}, infrared \cite{Bernhard1} and superfluid density \cite{Bernhard} measurements, then there should be radical changes in the superconducting state when the temperature falls below $T^* < T_c$. On cooling below $T_{\textrm{c}}$ one would expect the condensation free energy, critical fields and superfluid density to initially grow as if there were no pseudogap, consistent with the strong superconductivity seen in the overdoped region. Then, on crossing the $T^*$ line the mooted opening of the pseudogap would deplete this spectral weight such that these thermodynamic parameters would grow much more slowly on further cooling and perhaps even reduce in magnitude. Indeed, if there is a mean-field phase transition at $T^*$ then the $T$-dependent slope of these properties will change discontinuously. We refer to this general behaviour as `entrant' where the slope of $T^*(p)$ below $T_{\textrm{c}}$ remains negative as depicted in the inset to Fig.~\ref{freeen}(b). In the following we present a search for such behaviour.

\begin{figure}
\centering
\includegraphics[width=65mm]{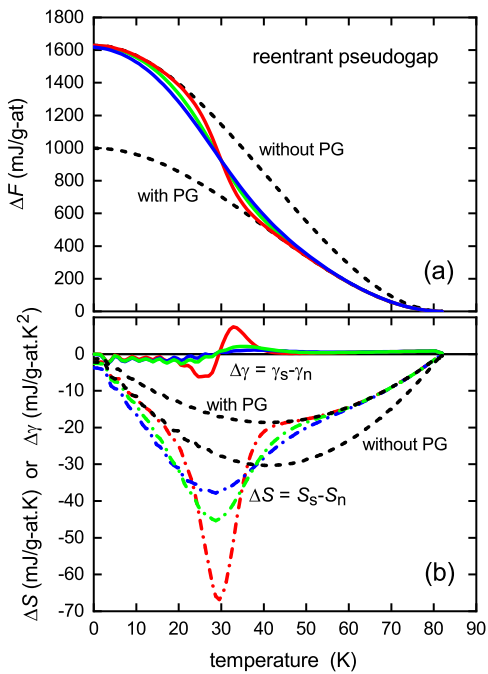}
\caption{\small
(Color online) The anticipated thermodynamic effects of reentrant crossover into the pseudogap on cooling in the superconducting state (a) condensation free energy, and (b) entropy, $\Delta S(T)$ (dash-dot curves), and specific heat coefficient, $\Delta \gamma(T)$ (solid curves). The dashed curves are the weak-coupling $d$-wave behavior, with $\Delta F(0)$ reduced by the pseudogap. The crossover is implemented using a $\tanh((T-30)/T_0)$ crossover function with $T_0$ = 5, 10 and 15 K as shown by the red, green and blue curves respectively. In such a crossover $\Delta S(T)$ is characterised by two inflexion points and $\Delta \gamma(T)$ by an additional large anomaly near 30 K.}
\label{bimodal}
\end{figure}

Further, we also test the converse of this, namely, if the slope of $T^*(p)$ below $T_{\textrm{c}}$ were positive as depicted by the gray line in Fig.~\ref{freeen}(a). We refer to this as `reentrant' behaviour. The motivation for this scenario is found in the ARPES study by Vishik {\it et al.} \cite{Vishik}. These authors claim to observe the pseudogap out to $p \approx 0.23$ just above $T_{\textrm{c}}$. $T^*(p)$ is then suggested to `back-bend', adopting a positive slope and falling to zero at $p^* = 0.19$. (This proposed back-bending phase diagram is also invoked by He {\it et al.}, \cite{He2}). If this were the case, then between $0.19 < p < 0.23$, cooling below $T_{\textrm{c}}$ means that the initial onset of superconductivity just below $T_{\textrm{c}}$ is already within the pseudogap state and all measures (critical fields, condensation energy and superfluid density) should then indicate weak superconductivity. However, on crossing the reentrant $T^*(p)$ line the superconductor will exit the pseudogap state into a strong superconductivity regime in which these measures are no longer suppressed and they will grow much more rapidly with further decreasing temperature. Such reentrant behaviour has both theoretical support \cite{Sachdev} and is experimentally observed in Ba(Fe$_{1-x}$Co$_x$)$_2$As$_2$ \cite{Nandi}.

The anticipated thermodynamic behavior is illustrated in Fig.~\ref{bimodal}. Panel (a) shows our calculation (black dashed curve) of the weak-coupling $d$-wave condensation free energy, $\Delta F (= \Delta F_{\textrm{ns}} = F_{\textrm{n}} - F_{\textrm{s}})$, as fitted to our experimental data for Bi$_2$Sr$_2$CaCu$_2$O$_{8+\delta}$ (Bi2212) just above $p^*$, i.e. without a pseudogap. (See Fig.~\ref{freeen} and also Won and Maki \cite{Maki}). The second black dashed curve is $\Delta F$ with the same temperature dependence but rescaled down in magnitude to give the same ground-state condensation energy observed in our Bi2212 samples at optimal doping, $p = 0.16$, where $T^* \approx T_c$, i.e. with a pseudogap present. We use a $\tanh((T-30)/T_0)$ crossover function between the two $\Delta F(T)$ curves with $T_0$ = 5, 10 and 15 K, as shown by the red, green and blue curves, respectively. These curves depict the effect of reentrant behavior on the condensation energy depending on how abrupt the crossover is. Literature examples, illustrative of this kind of behavior, include a low-temperature boost in superfluid density seen in Ba$_{1-x}$K$_x$Fe$_2$As$_2$ \cite{Khasanov}, and a boost in critical current density and superfluid density seen in PrOs$_4$Sb$_{12}$ \cite{Cichorek,Talantsev2}.

If such a boost does occur then Fig.~\ref{bimodal}(b) shows the effect on $\Delta S = -\partial\Delta F /\partial T$ (dash-dot curves) and $\Delta\gamma = \partial\Delta S/\partial T$ (solid curves), with red, green and blue having the same meaning as in panel (a). This illustrates quite generically a reasonably model-free expectation for a reentrant crossover, namely a double inflexion of $\Delta S(T)$ and an additional anomaly in $\Delta \gamma(T)$ below $T_{\textrm{c}}$ that can be very large depending on the narrowness of the crossover. The calculated $\Delta \gamma$ anomaly values of 7.4, 2.1 or 1.1 mJ/g.at.K$^2$ (for $T_0$ = 5, 10 or 15 K, respectively) are easily detectable by our differential measurements which are sensitive to $\pm 0.05$ mJ/g.at.K$^2$.

For an entrant pseudogap there will be a corresponding crossover from the upper dashed curve in Fig.~\ref{bimodal}(a) to the lower dashed curve, with consequent marked anomalies in $\Delta S$ and $\Delta\gamma$, including the possibility of a positive excursion in $\Delta S$ for a narrow crossover.

As noted, Vishik {\it et al.} \cite{Vishik} are not alone in proposing that the normal-state pseudogap persists to $p \approx 0.23$. Raman scattering in $B_{1g}$ symmetry has been interpreted to suggest that the pseudogap just above $T_{\textrm{c}}$ extends to $p \approx 0.23$, the putative location of the van Hove singularity, but not beyond \cite{Sacuto1}. Legros {\it et al.} \cite{Taillefer}, using high-field transport studies, promote the same picture. These would give additional apparent support to a reentrant phase diagram.

Either way, entrant or reentrant behaviour should give a complex, non-BCS-like $T$-dependence of critical fields, condensation energy and superfluid density providing either a downturn or a boost, respectively, to these properties on traversing the $T^*$ line. If there is a mean-field thermodynamic transition at $T^*$ then these changes will be abrupt. Moreover, as the pseudogap is effective in causing a large reduction in electronic entropy \cite{Loram} these effects should be substantial. Here we report the complete absence of such anomalous thermodynamic features. This includes the absence of inflexion points in the electronic entropy, in the entropy difference between normal and superconducting states and in the entropy change in applied field. Neither do we observe any anomaly in the various free energies investigated including the field-dependent free energy.

In Appendix B we summarise the extensive evidence from many different spectroscopies and transport measurements showing that the pseudogap closes at $p \approx 0.19$. Appendix C briefly argues that this result is also applicable to a wider group of what we call {\it canonical} cuprates.

\section*{2. Experimental details}

The ARPES \cite{Vishik}, Raman \cite{Sacuto1} and high-field transport \cite{Taillefer} studies, whose conclusions we dispute, all concern the generic system Bi$_2$Sr$_2$CaCu$_2$O$_{8+\delta}$ (Bi2212). As a consequence we focus solely on this system so that we are discussing precisely the same materials. However, we note that almost identical results were obtained for Y$_{0.8}$Ca$_{0.2}$Ba$_2$Cu$_3$O$_{7-\delta}$ (Y-123) which, like Bi2212, can also be overdoped.

Differential specific heat measurements on Bi2212 samples were originally made in the period 1998-2000 and the basic results were reported in two publications \cite{Loram,Loram4}. However, we present here additional details, analysis and insights. We synthesised and investigated three distinct polycrystalline samples: Bi$_{2.1}$Sr$_{1.9}$CaCu$_2$O$_{8+\delta}$, Bi$_{2.1}$Sr$_{1.9}$Ca$_{0.7}$Y$_{0.3}$Cu$_2$O$_{8+\delta}$ and Bi$_{1.8}$Pb$_{0.3}$Sr$_{1.9}$CaCu$_2$O$_{8+\delta}$ where the second material allows lower doping and the third allows higher hole doping than the parent material. The 0.1 excess Bi is found to reside on the Sr site \cite{Nature} and is necessary to achieve single-phase materials. The samples were synthesised by repeated solid-state reaction in stoichiometric quantities and each, including the reference, was approximately 0.9 g in weight.  The samples were fully oxygenated (overdoped) then subjected to a series of anneals progressively lowering the oxygen content and hence the doping state. Temperature- and field-dependent specific heat measurements were carried out for each successive doping level.

As described previously \cite{Loram4}, the measurements use a high-resolution differential technique where the specific heat difference between the sample and a reference is measured to a precision of 1 part in $10^5$. The reference was chosen to be 4\% cobalt-doped Bi$_{2.1}$Sr$_{1.9}$CaCu$_2$O$_{8+\delta}$, where the Co substitutes on the Cu sites, reducing $T_{\textrm{c}}$ rapidly due to strong scattering in combination with a nodal $d$-wave order parameter. While in other cuprates the chosen reference has been a Zn substituted sample \cite{Loram} the solubility of Zn in Bi2212 is low, while that of Co is up to at least 10\% \cite{Tallon5}. With 4\% Co substitution $T_{\textrm{c}}$ can be reduced to zero by deoxygenation so that there is no superconducting anomaly observed in the specific heat coefficient of the reference sample.

The mass of the reference is chosen so as to have as close as possible the same number of atoms as the sample. The differential technique thus allows most of the phonon term to be backed off enabling the much smaller electronic specific heat to be separated from the lattice term \cite{Loram2}. This still leaves a small residual phonon term which peaks at 16.5 K but this is found to scale precisely with the change in oxygen content of the sample and so can be identified and eliminated \cite{Loram4}. Then, because a sequence of 11 doping states were studied per sample the deduced electronic specific heat coefficient, $\gamma(T)$, is differenced relative to the end doping state thus automatically removing any possible residual contribution from the reference.

Most significantly, it was found that, for the same doping state, the three Bi2212 compositions yielded essentially identical thermodynamic parameters (see Fig. 3 of \cite{Loram}) so we conclude that the thermodynamic properties are independent of cation cross-substitution (and associated minor disorder), they are essentially dependent only on doping and thus represent the generic thermodynamic properties of this system. The extracted data is thus applicable to, and directly comparable with, the various Bi2212 samples used in the above-noted ARPES, Raman and high-field transport measurements. The calculation of various thermodynamic functions from the data, including the superfluid density and the relationship between electronic entropy and spin susceptibility is described in Appendix A. Henceforth, all references to thermodynamic functions concern the electronic term only and so we now drop the descriptor ``electronic".

Finally, for consistency of comparison we evaluate doping, $p$, in the same manner as used in the other spectroscopic studies which we discuss, namely from the ratio of $T_c/T_c^{\textrm{max}}$ by inverting the empirical parabolic phase curve $T_c = T_c^{\textrm{max}}[1 - 82.6(p-0.16)^2]$ \cite{Presland}. In addition, we also measured the room temperature thermopower to determine $p$ \cite{Obertelli} and this gave completely consistent results. Further, the residual phonon term in $\gamma(T)$ scales precisely with changing oxygen content thus enabling very precise measures of increments in $\delta$ in the chemical formulae. As shown previously\cite{Presland,Anukool}, we find $\Delta\delta = \Delta p$ as expected if the two doped holes per additional oxygen are shared between the two CuO$_2$ planes per formula unit.

Using these characterisations we may be more precise in our definition of the pseudogap closing point, described above as $p^* \approx 0.19$, by the rather narrow range $p^* = 0.19 \pm 0.005$.

\begin{figure}
\centering
\includegraphics[width=70mm]{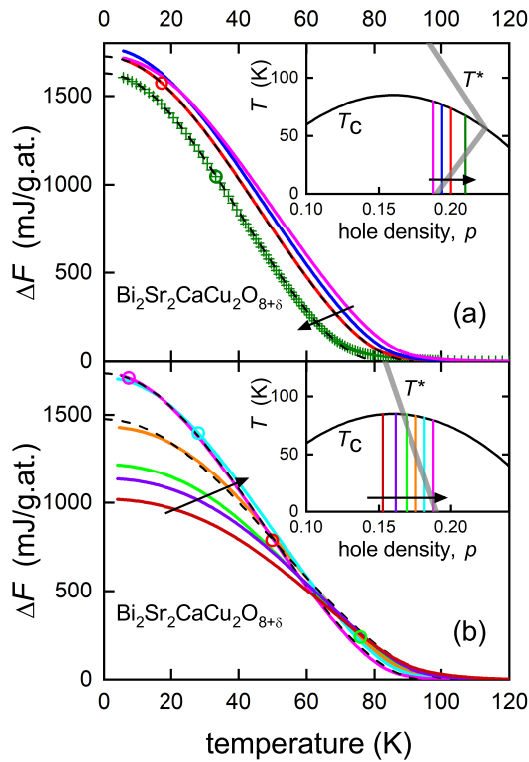}
\caption{\small
The two $T^*(p)$ scenarios tested in this work (grey lines in insets): (a) `reentrant' where the putative pseudogap line meets the $T_c(p)$ phase curve then back-bends to fall to zero at critical doping $p^* = 0.19$ ; and (b) `entrant' where the putative pseudogap line $T^*(p)$ falls monotonically to zero at $p^* = 0.19$ with positive slope. Colored curves in the main panels show the measured condensation free energy, $\Delta F(T)$, obtained by integrating the electronic condensation entropy, $\Delta S(T)$. The curves in each main panel are at doping levels indicated by the colour-coded vertical lines on the phase diagrams shown in the insets. The excellent agreement with near-weak-coupling $d$-wave BCS free energy calculations (four dashed curves) rule out both the reentrant and entrant scenarios. Small deviations near $T_{\textrm{c}}$ are due to superconducting fluctuations.
}
\label{freeen}
\end{figure}

\section*{3. Results and analysis}
\subsection*{3.1 Condensation free energy}

Fig.~\ref{freeen} shows the condensation free energy, $\Delta F(T) = \Delta F_{\textrm{ns}}$, calculated from the experimental data by integrating $\Delta S_{\textrm{ns}} = S_{\textrm{n}} - S_{\textrm{s}}^{\textrm{expt}}$ for nine different doping states. As described in Appendix A.2 the normal-state entropy $S_{\textrm{n}}$ (and $\gamma_{\textrm{n}}$) are calculated from a rigid ARPES-derived dispersion with our previously-reported pseudogap model \cite{Storey4} including ungapped Fermi arcs. (Note that the $\Delta F(T)$ data reported elsewhere \cite{Tallon2} were for a nodal pseudogap model. The impact of the two differing models on $\Delta F(T)$ is compared in Appendix D).

The magenta curves in panels (a) and (b) are at critical doping $p = p^* = 0.188$ while (a) shows three other curves for higher doping ($p$ = 0.194, 0.20, 0.21) which straddle the putative reentrant $T^*(p)$ line of Vishik {\it et al.} \cite{Vishik}, and (b) shows five other curves for lower doping ($p$ = 0.182, 0.176, 0.169, 0.162, 0.153) which straddle the putative entrant $T^*(p)$ line reported, for example, by Zaki {\it et al.} \cite{Zaki} or Naqib {\it et al.} \cite{Naqib}. The data curves are plotted as linear segments (rather than smooth spline curves) but the quality of the data is illustrated in panel (a) for the most heavily doped sample (olive green curve) where just every fourth data point is plotted using the green crosses.

The black dashed curves in both panels are our calculated mean-field near-weak-coupling $d$-wave temperature dependence of $\Delta F$. The data is fitted simply by setting the magnitude of the mean-field transition temperature, $T_c^{\textrm{mf}}$, and the value of the ground-state condensation energy, $\Delta F(0) \equiv \Delta U(0)$. We used the Padamsee $\alpha$-model \cite{Padamsee} recalculated for $d$-wave superconductivity with the value $2\alpha = 2\Delta/(k_BT_c^{\textrm{mf}}) = 4.5$, as inferred previously \cite{Tallon3,Tallon2}. (Note that the weak-coupling value is 4.288 \cite{Maki}). There is an excellent match between the simple mean-field behaviour and the observed condensation free energy across the entire $T$-range except close to $T_{\textrm{c}}$ where superconducting fluctuations are the cause of the small discrepancies. In the underdoped samples the discrepancy due to fluctuations is larger, mainly due to the fact that $T_c^{\textrm{mf}}$ is so much greater than $T_{\textrm{c}}$ here, but also because our Padamsee model does not include an antinodal pseudogap which also affects the $T$-dependence near $T_{\textrm{c}}$. However, it is important to note that these small fluctuation-induced deviations seen here in the underdoped region below $p_{\textrm{crit}}$ are the opposite to what would be expected in the entrant scenario. On cooling towards and below $T_{\textrm{c}}$ we see that $\Delta F(T)$ actually rises more slowly at first (due to fluctuations) then quickly develops its full weight, with the $T$-dependence over most of the range below $T_{\textrm{c}}$ following the simple mean-field behavior.

Lastly, the color-coded circles on each curve in both panels show where the putative $T^*$ is expected from the two insert figures. Notably, there are no knees or kinks observed in $\Delta F(T)$ at these points.

The excellent match between mean-field behaviour and the observed condensation energy is significant in light of the mooted reentrant (a) or entrant (b) behaviour. In the former case $\Delta F(T)$ should rise more slowly at first as though heading for a small ground-state value (reduced due to the pseudogap) then abruptly upturn at $T^*$ (see circles) as the superconductor moves out of the reentrant pseudogap state. Instead all of these overdoped samples follow the canonical behaviour for a single order parameter, all exhibiting strong superconductivity with a similar ground-state condensation energy. Indeed it has been shown that, across this overdoped region, the BCS ratio $\Delta F(0)/(\gamma_n\,k_B\,T_{\textrm{c,mf}}^2)$ adopts a constant value of $\approx 0.17$, as expected for near weak-coupling $d$-wave superconductivity \cite{Tallon2}. The small decrease in $\Delta F(0)$ at the highest doping is therefore simply due to the fall in $T_{\textrm{c}}$ on the overdoped side. In contrast, panel (b) shows a rapid fall in $\Delta F(0)$ as doping falls below $p^*$. This is due to the abrupt opening of the pseudogap at $p^*$ removing antinodal states that would otherwise be available for superconductivity. It is clear that these states are completely removed at all temperatures below $T_{\textrm{c}}$ and not just at a putative $T^*$ (circles) below $T_{\textrm{c}}$. Despite this rapid fall in $\Delta F(0)$, each of the curves rises monotonically, free of any semblance of a knee and consistent with a single-order-parameter mean-field behaviour as shown by the dashed curves.

\begin{figure}
\centering
\includegraphics[width=70mm]{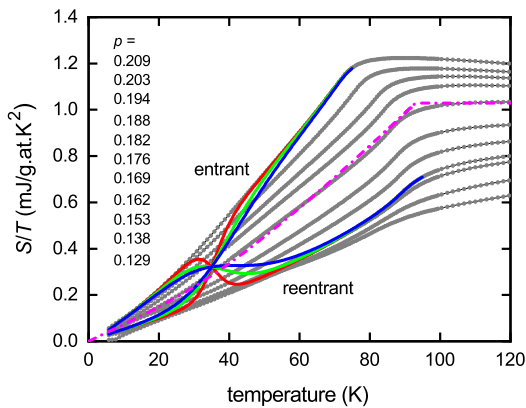}
\caption{\small
(Color online) The temperature dependence of $S/T$ for pure Bi2212. Gray data points: experimental entropy data (every fourth data point shown). Magenta dash/dot curve: our calculation of $S/T$ using the Padamsee $\alpha$-model for a $d$-wave order parameter in the near-weak-coupling case where $2\Delta/k_BT_c = 4.5$. The expected entrant or reentrant crossover behavior into, and out of, the pseudogapped state is shown by the red, green and blue curves. The crossover is implemented using a $\tanh((T-35)/T_0)$ function with $T_0$ = 5, 10 and 15 K as shown by the red, green and blue curves respectively. No such crossover is evident, even weakly so, in the experimental data.
}
\label{rawentropy}
\end{figure}

Crucially, this conclusion is independent of the details of the normal-state entropy fits, noted above, which are necessary to construct the condensation energy. To see this we show in Fig.~\ref{rawentropy} the {\it as-measured} entropy data from Loram {\it et al.} \cite{Loram}, plotted as $S(T)/T$, for the full eleven doping states measured for pure Bi2212 (gray data points). This is obtained by integrating the experimental $\gamma(T)$ data. For clarity, the plotted data show every fourth data point. It is evident that the measured data is free of any obvious entrant or reentrant anomaly. The magenta dash/dot curve shows our calculation of $S(T)/T$ for a $d$-wave order parameter where $2\Delta/k_BT_c = 4.5$ i.e. near weak coupling, as calculated using the Padamsee $\alpha$-model \cite{Padamsee}. Evidently, any deviation in the experimental data from this canonical behavior is very small except near $T_{\textrm{c}}$ where fluctuations cause a rounding of the transition.  Also plotted is the expected entrant or reentrant behaviour if there should occur a crossover into, or out of, the pseudogap state. Again we have used crossover ranges of $T_0$ = 5, 10 and 15 K (red, green and blue curves). There is no suggestion in the data of even a weak crossover into, or out of, the pseudogap state. For the optimally doped sample (the end-point of the crossover) we infer $E^*$ = 16 meV. A crossover anomaly corresponding to a pseudogap 1/10$^{\textrm{th}}$ this size would probably be observable in our data. Thus we suggest that if there is an entrant or reentrant pseudogap near $p^*$ its magnitude is less than 2 meV.

\begin{figure}
\centering
\includegraphics[width=70mm]{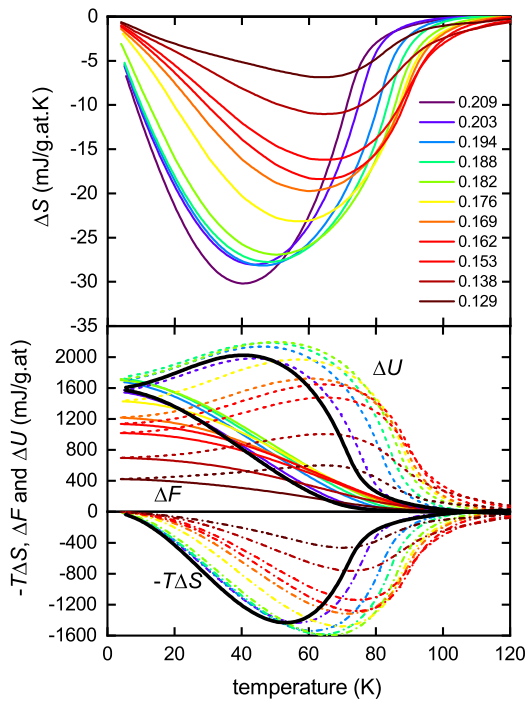}
\caption{\small
(a) The $T$-dependence of the electronic condensation entropy $\Delta S = S_s - S_n$ for Bi2212 at the various doping levels shown, from $p$ = 0.162 to 0.209. All curves are totally free of the inflexions shown in Fig.~\ref{bimodal} characteristic of reentrant behavior. The curvature near $T_{\textrm{c}}$ arises from strong superconducting fluctuations. (b) The $T$-dependence of $\Delta F$ and its component terms $\Delta U$ and $-T\Delta S$. The color coding is the same as in panel (a). Notably the strong fluctuation terms seen in both $\Delta U$ and $T\Delta S$ are almost completely absent in $\Delta F$ - where the respective curves for $p=0.209$ are highlighted for clarity.
}
\label{internalen}
\end{figure}

We note in passing that the low-$T$ slope of $S/T$ is proportional to $\Delta_0^{-1}$. The convergence of the four underdoped curves at low temperature shows that the gap magnitude saturates at low doping while, with increasing doping, the gap magnitude is seen to fall increasingly out to the highest overdoped state. A similar result is reported in ARPES measurements \cite{Vishik}.

\begin{figure}
\centering
\includegraphics[width=70mm]{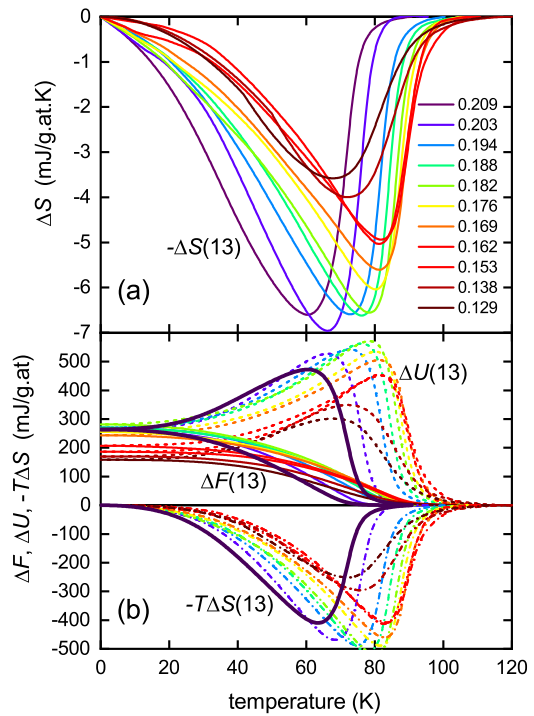}
\caption{\small
(Color online) (a) The $T$-dependence of the magnetic entropy difference, $\Delta S(13) = S(13) - S(0)$, obtained in changing field from 0 to 13 tesla, obtained by integrating the $\Delta\gamma(13) = \gamma(13) - \gamma(0)$ data for Bi2212 shown in Fig.~\ref{chi}(c). Doping levels are given in the legend. (b) the magnetic free energy difference $\Delta F(13)$ and its internal energy and entropy components obtained by integrating the same $\Delta\gamma(13)$ data using Eq.~\ref{free-energy}. As in Fig.~\ref{internalen} the curvature near $T_{\textrm{c}}$ arises from strong superconducting fluctuations, though less marked here. The color coding is the same as in panel (a). As in Fig.~\ref{internalen}(b) the strong fluctuations seen in both $\Delta U(13)$ and $T\Delta S(13)$ are almost completely absent in $\Delta F(13)$. The respective curves for $p=0.218$ are highlighted for clarity.
}
\label{magentropy}
\end{figure}

\subsection*{3.2 Condensation entropy}
The benefit of investigating $\Delta F_{\textrm{ns}}(T)$ for entrant or reentrant behavior is that it is otherwise a simple monotonic function. The entropy $\Delta S(T)$ is non-monotonic but it is, as shown in Fig.~\ref{bimodal}, more sensitive to anomalous behavior because of the derivative (and $\Delta \gamma(T)$ even more so). Fig.~\ref{internalen}(a) shows the condensation entropy $\Delta S(T) = S_{\textrm{n}}(T) - S_{\textrm{s}}(T)$ for Bi2212 where, as noted, $S_{\textrm{n}}(T)$ is calculated from the ARPES derived dispersion as described in Appendix A. The entropy extrapolates to zero at $T=0$, as required, and the curvature extending well above $T_{\textrm{c}}$ arises from strong superconducting fluctuations \cite{Tallon3}. (These are strong in the sense that the ratio $T_{\textrm{c}}^{\textrm{mf}}/T_{\textrm{c}}$ can be as large as 1.6. Note that this conclusion does depend on correctly identifying the mean-field behavior of $\gamma(T)$ below the fluctuation region below $T_{\textrm{c}}$ \cite{Tallon2,Tallon3}.) Apart from these features the entropy is totally devoid of any inflexion or flattening as might be indicative of reentrant (refer Fig.~\ref{bimodal}) or entrant behavior, respectively. The condensation free energy is shown in panel (b) along with its components $\Delta U(T)$ and $T\Delta S(T)$. Given the form of Eq.~\ref{free-energy} it is clear that for any reasonably sharp anomaly in $\gamma(T)$ (whose width is less than 10-20\% of $T_{\textrm{c}}$) there will always be significant cancellation in $\Delta U - T\Delta S$. But note that the non-monotonic behavior of each of these components exactly cancel so as to leave a simple monotonic mean-field behavior in $\Delta F(T)$. Note also that the strong superconducting fluctuations evident in these component terms largely offset each other so as to almost totally suppress the fluctuation term in the free energy. This is an important warning that some thermodynamic and spectroscopic features are more sensitive to fluctuations than others. Those that are very sensitive to fluctuations (like the entropy) include the spin susceptibility (see Appendix A and its consequent relevance to NMR studies), the superconducting gap function which extends well above $T_{\textrm{c}}$ and is proportional to $2\Delta F(T) + T\Delta S(T)$ \cite{Tallon2}, and the Raman scattering cross-section \cite{Storey1}. As a consequence it is easy to confuse a normal-state partial gap arising from superconducting fluctuations with the pseudogap and it is necessary to additionally use magnetic fields to suppress one and not the other in order to definitively distinguish them \cite{Naqib,Alloul2}.

\subsection*{3.3 Magnetic entropy}
Fig.~\ref{chi}(c) shows the $T$-dependence of $\Delta \gamma(13,T) = \gamma(13,T) - \gamma(0,T)$, the change in $\gamma$ due to increasing field from 0 to 13 tesla. (This will be discussed later relative to the other panels in the figure with respect to fluctuations). This data is integrated with respect to temperature, as in Eq.~\ref{free-energy} of Appendix A, to obtain the field-dependent free energy difference, $\Delta F(13,T)$, and its internal energy and entropy components. These are plotted in Fig.~\ref{magentropy}. $\Delta F(13,T)$ is related to the superfluid density, $\rho_s(T) = \lambda^{-2}(T)$, as detailed in Appendix A.3. Considering $\Delta S(13,T)$ first, we note that the inflexion associated with putative reentrant behaviour, as illustrated in Fig.~\ref{bimodal}(b), is absent on the high-temperature side of the extremum. There is a significant inflexion on the low-temperature side but this is expected under the London model, and is present in all samples. The strong superconducting fluctuation feature seen in $\Delta U(13,T)$ and $-T\Delta S(13,T)$ is again nearly completely cancelled out in $\Delta F(13,T)$.

%\begin{figure}
%\centering
%\includegraphics[width=70mm]{Figure6.eps}
%\caption{\small
%Normalised plot of $\Delta F(13,T,p)/\Delta F(0,T,p)$ versus $T/T_c$ for 11 doping states (annotated) for Bi2212. None reveal any features that might be associated with entrant or reentrant behavior in the closing or opening of the pseudogap.
%}
%\label{DF_scaled}
%\end{figure}

A normalised plot of $\Delta F(13,T,p)/\Delta F(0,T,p)$ versus $T/T_c$ reveals that all doping states have very similar temperature dependences. None shows a downturn or a boost relative to the others that might reflect entrant or reentrant behavior. These are not expected to have identical $T$-dependences. Both the opening of the antinodal pseudogap and the London model for $\Delta F(13,T,p)$ will produce small systematic variations in the scaled behavior depending e.g. on the amplitude of the superfluid density. But there is no substantial deviation that would signal the opening or closing of the pseudogap below $T_{\textrm{c}}$.

\begin{figure}
\centering
\includegraphics[width=70mm]{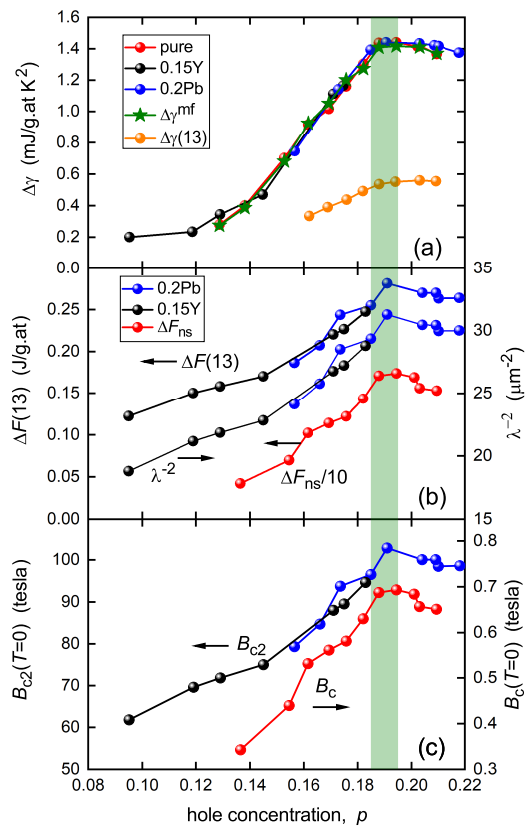}
\caption{\small
(Color online) Summary of the doping dependence of various thermodynamic quantities: (a) shows the jump, $\Delta\gamma(T_c,p)$, in  specific heat coefficient at $T_{\textrm{c}}$ for pure Bi2212, 0.15 Y-substituted Bi2212 and 0.2Pb substituted Bi2212. Also shown is the previously-reported true mean-field jump, $\Delta\gamma^{\textrm mf}(p) = \Delta\gamma(T_c^{\textrm mf},p)$, deduced from an entropy balance analysis of the fluctuation specific heat \cite{Tallon3}; and the difference in jump height, $\Delta \gamma(13,p)$ between 0 and 13 tesla. (b) shows the ground-state amplitude of the field-dependent change in free energy, $\Delta F(13,T=0,p)$ and $\Delta F_{\textrm{ns}}$ in zero field; and the superfluid density, $\lambda_0^{-2}(p)$ deduced from Eq.~\ref{HaoClem}. (c) shows $B_{\textrm{c2}}(0)$ determined from $\Delta F(13,0)$ and $B_{\textrm{c}}(0)$ from $\Delta F_{\textrm{ns}}(0)$.
}
\label{gamma_jump}
\end{figure}

Fig.~\ref{gamma_jump} summarises the doping dependence of a number of key thermodynamic amplitudes. Panel (a) shows the previously-reported jump, $\Delta\gamma(T_{\textrm{c}},p)$, in  specific heat coefficient at $T_{\textrm{c}}$ for pure Bi2212, 0.15 Y-substituted Bi2212 and 0.2Pb substituted Bi2212 \cite{Loram}. We earlier showed that this jump comprised an apparent mean-field step and a fluctuation step. Using an entropy balance analysis of the fluctuation specific heat we were able to deduce the `true' mean-field step, $\Delta\gamma^{\textrm mf}(p) = \Delta\gamma(T_{\textrm{c}}^{\textrm mf},p)$, that would have occurred at the significantly higher $T_{\textrm{c}}^{\textrm mf}$ if fluctuations were suppressed \cite{Tallon3}. $\Delta\gamma^{\textrm{mf}}$ is plotted in the figure by the green stars. These track the observed total $\Delta\gamma$ values rather closely.  Also shown is the difference in jump height, $\Delta \gamma(13,p)$, at 0 and 13 tesla (gold symbols). The vertical green-shaded band shows the interval $p = 0.190 \pm 0.005$ around critical doping. These quantities all reveal an abrupt fall in amplitude beginning at $p\approx0.19$ with the opening of the pseudogap. Note that these include properties evaluated at $T_{\textrm{c}}$ or at $T_c^{\textrm mf}$ which is up to 40 K higher. While $\Delta\gamma(T_c,p)$ is not necessarily a true measure of the weight of the jump (and hence of the density of Cooper pairs) because transition widths may (and do) vary with doping, the deduced true mean-field step $\Delta\gamma^{\textrm mf}(p)$ is an integrated measure and thus largely independent of any variations in transition width. These data collectively reveal a sudden collapse of pair density on reducing the doping below $p \approx 0.19$ clearly signalling the opening of the pseudogap there.

Panel (b) of Fig.~\ref{gamma_jump} summarises the doping dependence of the {\it ground-state} values of the condensation free energy, $\Delta F_{\textrm{ns}} = F_{\textrm{n}} - F_{\textrm{s}}$ and the field-induced free energy change $\Delta F(13)$. These are $T=0$ properties and they both also show an abrupt fall on reducing the doping below $p \approx 0.19$ simultaneous with the abrupt fall in properties at $T_{\textrm{c}}$ and $T_c^\textrm{mf}$. The former, $\Delta F_{\textrm{ns}}$, defines the thermodynamic critical field, $B_{\textrm{c}} = \phi_0/[2\sqrt{2}\pi\xi\lambda]$, while the latter, $\Delta F(13)$, defines the ground-state superfluid density, $\lambda^{-2}(0)$, using the London model, Eq.~\ref{HaoClem}. (The detailed calculation of the full $T$-dependence of $\lambda^{-2}(T)$ is described elsewhere).  These parameters are all plotted in panels (b) and (c). Finally, having determined $B_{\textrm{c}}(0)$ and $\lambda^{-2}(0)$ we calculate $B_{\textrm{c2}}(0)$ which is plotted in panel (c). The peak value of this parameter at $p^*$ of 103 tesla is very reasonable and the implied value of $\xi(0) = 1.79$ nm consistent with various reported measurements \cite{Poole}.

It is evident from Fig.~\ref{gamma_jump} that all key parameters associated with the strength of the condensate commence their pseudogap-induced collapse at $p \approx 0.19$ whether at $T=0$, $T=T_c$ or $T=T_c^{\textrm mf}$ - a range of some 140 K. The onset of the pseudogap does not therefore follow a doping-dependent $T^*(p)$ line but is fixed at $p=p_\textrm{crit} \approx 0.19$, {\it independent of temperature}. Combining this with the repeated observation that the normal-state entropy above $T_{\textrm{c}}$ never recovers from lost states up to 300 K and higher \cite{Loram,Loram1,Loram3} we may conclude that the opening of the pseudogap occurs only at $p=p_\textrm{crit} \approx 0.19$ independent of temperature from $T=0$ to well over 300 K. This is consistent with our claim that the pseudogap line is an energy scale not a temperature scale \cite{Tallon1,Williams}. The inference is that the very weak but abrupt mean-field transitions observed near a monotonically descending $T^*(p)$ line in the underdoped regime occur {\it within} the preexisting pseudogap state which extends to very high temperatures. They are not transitions {\it into} the pseudogap state, and possibly not even associated with the pseudogap. The pseudogap may simply create conditions conducive to this correlation. Thus, for example, a pseudogap-induced change of the Fermi surface can allow nesting $q$-vectors that induce charge ordering \cite{Comin}.

\section*{4. Discussion}
\subsection*{4.1 Fluctuations.}
Despite our extensive results showing that the pseudogap closes at $p^* \approx 0.19$, independent of temperature, and the large amount of thermodynamic and spectroscopic data that supports this (see Appendix B), there persists the lingering view that there truly exists a pseudogap $T^*(p)$ line that extends deep into the overdoped region meeting the superconducting phase curve $T_c(p)$ around $p = p^* \approx 0.23$ \cite{Vishik,He2,Sacuto1,Taillefer} and projecting to zero just where $T_c \rightarrow 0$. How can two such different pictures still persist in such a mature field? We suggest that the confusion has two primary causes. Firstly, the existence of a partial gap in the DOS arising from superconducting fluctuations, and extending well above $T_{\textrm{c}}$, is easily confused with the pseudogap. And secondly, the literature actually confuses a spectroscopically-determined energy scale, $E^*(p)$, with a temperature scale, $E^*(p)/k_B$. In this section we discuss this situation in some detail.

%As a useful summary we collect together the evidence for the abrupt closure of the pseudogap at $p = p^* \approx 0.19$ in Appendix B.

%Now we address more closely the value of $p^*$ where the pseudogap closes. For Bi2212 it was established long ago by examination of many physical properties that $p^*=0.19$ holes/planar Cu \cite{Tstar}. The data is summarised in Loram {\it et al.} \cite{Loram}, in Fig. 1 of Tallon {\it et al.} \cite{TallonMOS} and Fig. 1 of Storey {\it et al.} \cite{Storey5}. Despite this robust determination there remains a widespread view, promoted most recently by Legros {\it et al.} \cite{Taillefer}, that this $T^*(p)$ line extends deep into the overdoped region, joining the $T_c(p)$ phase curve projecting to zero just where $T_c \rightarrow 0$.

In many cases nominal $T^*$ values are obtained by identifying `kinks' or downturns in the $T$-dependence of various physical properties. The problem is that above $T_{\textrm{c}}$ one must take superconducting fluctuations into account. Experimentally these are seen most easily from a field-dependent downturn in the in-plane electrical resistivity and a diamagnetic contribution to the static susceptibility which is larger for magnetic fields perpendicular to the CuO$_2$ planes.  It is not as widely recognised that in overdoped Bi2212 superconducting fluctuations can also reduce the electronic density of states (DOS) at the Fermi level  \cite{Larkin,Watanabe2,Benseman} and hence decrease the spin susceptibility and affect all other properties which depend on the DOS.  Although we are not aware of an explicit theoretical treatment, in our work a decrease in the measured electronic entropy, as $T_{\textrm{c}}$ is approached from above, must mean that there is a decrease in the electronic DOS. This must be true because there is no other source of entropy once the phonon part has been correctly subtracted.

\begin{figure}
\centering
\includegraphics[width=65mm]{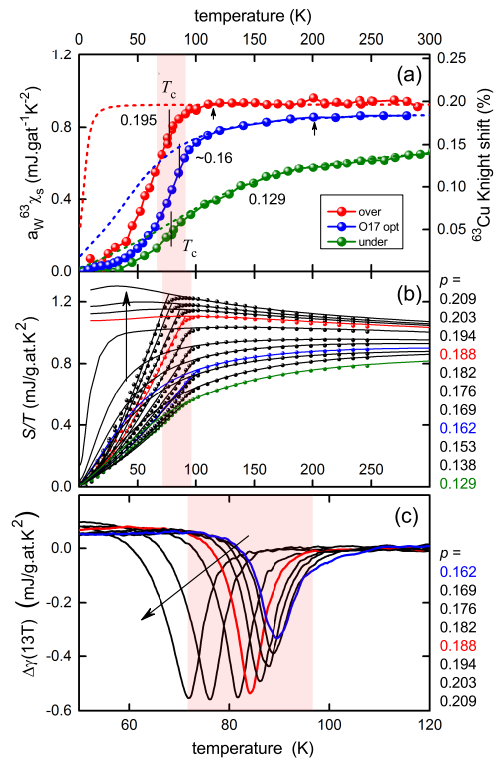}
\caption{\small
(a) $^{63}$Cu Knight shift for Bi2212 \cite{Ishida} expressed in \% shift and in entropy units using $a_W \chi_s$; doping, $p$, is annotated. The $^{17}$O shift for $p$=0.16 is reproduced from \cite{Takigawa}. Dashed curves: normal-state pseudogap fits with $E^*$ = 1.4, 13.4 and 27.6 meV. (b) Electronic entropy coefficient $S/T$ for 11 doping states from \cite{Loram}. Curves for critical doping where the pseudogap closes, optimal doping and most underdoped are shown by red, blue and green, respectively, and correspond to the color coding in (a). All curves are calculated \cite{Storey4} using the full ARPES-derived dispersion involving Fermi arcs as applicable to both normal and superconducting states, while those in Fig. 1 use the non-weak-coupling Padamsee calculation for $\Delta F(T)$ (see text). (c) The change, $\Delta \gamma(13) = \gamma(13)-\gamma(0)$, in specific heat coefficient, $\gamma$, between 0 and 13 T for doping levels listed at right.
}
\label{chi}
\end{figure}

The point is illustrated in some detail in Fig.~\ref{chi}(a) where the $^{63}$Cu Knight shift data of Ishida {\it et al.} \cite{Ishida} is reproduced and compared with the thermodynamic functions in panels (b) and (c). The intimate relationship between the spin susceptibility and entropy is detailed in Appendix A.4. This NMR data was used by Legros {\it et al.} \cite{Taillefer} to augment their `phase diagram' (see their Supplementary Information) which was largely adapted from Vishik {\it et al.} \cite{Vishik}. Also shown is the $^{17}$O Knight shift data from Ishida's Fig. 3 as earlier reported by Takigawa and Mitzi \cite{Takigawa}. Spin shifts are thus presented for overdoped ($p=0.195$), optimal doped ($p \approx 0.16$) and underdoped ($p=0.129$), as annotated. The scale on the RHS shows the Knight shift referenced to the $T$-independent orbital shift. Note that a spin shift of $^{17}$K = 0.2 (\%) corresponds to $^{63}$K = 0.177 (\%). The small vertical arrows show the $T^*$ values where Ishida {\it et al.} identified the opening of the pseudogap (in our opinion erroneously). The dashed curves are our pseudogap fits to their normal-state data. As before \cite{Tallon3}, we have assumed an otherwise flat band with a pseudogap represented by a simple V-shaped electronic DOS centered on $E_F$ and remaining finite there, thus reflecting the presence of residual Fermi arcs. The susceptibility is calculated using the standard formula for weakly-interacting Fermions:
\begin{equation}
\chi_s = -2\mu_B^2 \int_{-\infty}^\infty \! \frac{\partial f}{\partial E} \, N(E) \,\mathrm{d}E
\label{entropy}
\end{equation}
\noindent where $f(E)$ is the Fermi function. This, with the V-shaped DOS, gives the analytical formula:

\begin{align}
\chi_s(T) &= 2N_0\mu_B^2  \nonumber \\
&\times\left[1 - \frac{N_0-\nu_0}{N_0}\left(\frac{E^*}{2k_BT}\right)^{-1} \ln\left[\cosh\left(\frac{E^*}{2k_BT}\right)\right]\right]
\label{chi_anal}
\end{align}
\noindent where $N_0$ is the background DOS, $\nu_0$ is the residual DOS at $E_F$ (arising from the Fermi arcs) and $E^*$ is the gap edge of the pseudogap V.

A more realistic model involving the ARPES-derived dispersion, as in the calculation \cite{Storey4} of the normal-state entropy in Fig.~\ref{chi}(b), gives almost identical results and does not alter our conclusions. In this model significant deviations only occur closer to the van Hove singularity (vHs), proximity to which is evidenced in the low-temperature upturn in $S/T$ in the most overdoped samples in Fig.~\ref{chi}(b). The three NMR data sets are far enough away from the vHs to not be affected.

The three dashed curves in Fig.~\ref{chi}(a) for over-, optimal- and under-doped Bi2212 are the fits with pseudogap energies of $E^*$ = 1.4, 13.4 and 27.6 meV, respectively. (In the first of these $E^*$ has probably already vanished at this doping but we retain the small value of $E^*$ = 1.4 meV to illustrate below that large curvature, as seen in this instance, can only occur at low temperature when $E^*$ is very small. So the large curvature observed in NMR data near 90 K \cite{Ishida} cannot be ascribed to the speudogap).

The LHS scale shows the spin susceptibility in entropy units, $a_W \chi_s$, obtained from multiplying $\chi_s$ by the Wilson ratio for weakly interacting Fermions, $a_W = (\pi k_B/\mu_B)^2/(3\mu_0)$ - see Appendix A.4.

As already noted, Fig.~\ref{chi}(b) shows the experimental \cite{Loram} and modelled \cite{Storey4} electronic entropy coefficient calculated for Bi2212 from the ARPES-derived dispersion \cite{Kaminski}. To better expose the detail this plot shows only every 20$^{\textrm{th}}$ experimental data point. Notably, the absolute magnitudes in panels Fig.~\ref{chi}(a) and (b) are very similar, underscoring the previous observation \cite{Loram1} that $S/T$ and $\chi_s$ are much the same in Fermi units, consistent with the elementary excitations being weakly interacting Fermions. Note that this is also true for many strongly correlated heavy Fermion systems \cite{Jaklic}.

As mentioned, the two small arrows in Fig.~\ref{chi}(a) indicate $T^*$ values inferred by Ishida {\it et al.} \cite{Ishida} for the optimal and overdoped samples. The two points are reproduced in the plot of $T^*(p)$ reported  by Legros {\it et al.} \cite{Taillefer} with the overdoped sample showing $T^* \approx 110$ K, well above $T_c \approx 79$ K. However, the normal-state fits (dashed curves) combined with the corresponding entropy data (red data and curves in panel Fig.~\ref{chi}(b)) show that the downturn just above $T_{\textrm{c}}$ is associated with superconducting fluctuations which reduce both the spin susceptibility and the measured entropy via their effect on the electronic DOS. It is unrelated to the pseudogap.  This is even more evident in the specific heat coefficient, $\gamma(T)$ where the fluctuation contribution is symmetric about $T_{\textrm{c}}$ \cite{Tallon2,Tallon3} - see Fig.~\ref{chi}(c). A lingering criticism of the electronic specific heat data is the question as to whether the differential technique really has succeeded in accurately backing-off the much larger phonon contribution. To meet this concern we show in Fig.~\ref{chi}(c) the change, $\Delta\gamma(H)$, in $\gamma$ between zero external field and $\mu_0 H=13$ T as reported by Loram {\it et al.} \cite{Loram}. This difference automatically eliminates any residual phonon contribution and we see in the field-dependent anomaly the effect of a magnetic field in suppressing fluctuations above and below $T_{\textrm{c}}$. The fluctuation range for the sample at critical doping (red curve) is shown by the pink shading and this fluctuation range is reproduced by the shading in Fig.~\ref{chi}(a) and (b). (Note that despite the same doping state the $T_{\textrm{c}}$ values are somewhat different for the single crystal in (a) with $T_c=79$ K and the polycrystal in (b) with $T_c=83.9$ K. Such differences between crystals and polycrystals are not uncommon. They probably arise from the presence of slightly more in-plane defects in state-of-the-art single crystals.) In summary, the similarities between the downturns in $^{63}K(T)$ and the measured entropy just above $T_{\textrm{c}}$, coinciding as they do with the pink-shaded fluctuation range, together with the evidence from tunnelling studies in overdoped Bi2212 \cite{Watanabe2,Benseman} lead us to conclude that these downturns arise not from the pseudogap but from superconducting fluctuations which reduce the electronic DOS at $E_F$, and that the pseudogap has already closed at this critical doping level or is very small, i.e. $E^*/k_B \approx 0$ (and certainly $T^* \neq 110$ K).

The vanishing of $E^*$ at this doping may not be immediately evident from the Knight shift but the benefit of placing $\chi_{\textrm{s}}$ alongside its complementary thermodynamic variable $S/T$ is that any small loss of normal-state electronic entropy at low T (due to a small residual pseudogap) must also be reflected in the actual measured entropy near $T_{\textrm{c}}$ due to entropy conservation. This is more easily seen in $\gamma = \partial S/\partial T$, as the area under a $\gamma(T)$ curve is entropy. Thus any (small) gapping in the DOS will produce a (small) reduction in the low-$T$ normal-state specific heat coefficient, $\gamma_{\textrm{n}}$, and therefore, by entropy conservation, the peak in specific heat at $T_{\textrm{c}}$ will also be reduced. Fig.~\ref{gamma_jump}(a) shows that the jump, $\Delta\gamma$, at $T_{\textrm{c}}$ remains constant in the overdoped region and then only below critical doping (red curve) does it begin to collapse, and it does so rapidly. In principle the same effect might be seen by differentiating the spin susceptibility, $\partial(\chi_{\textrm{s}}T)/\partial T \propto \chi$, but this would require very good data and there is no spin analogue of the third law of thermodynamics that leads to ``susceptibility conservation”. The pseudogap has closed at this critical doping state.

To pursue this in even more detail we consider the evolution of $S/T$ with doping shown in Fig.~\ref{chi}(b). For the most underdoped sample (green data and curves) the pseudogap is large, with $E^*/k_B = 322$ K, and $S/T$ is broadly curved over a comparable temperature range - see also the fits to the green NMR data points in Fig.~\ref{chi}(a). With increasing doping $E^*$ falls and the curvature increases, but as it does so the region of high curvature falls towards $T=0$, as also shown in the fits in Fig.~\ref{chi}(a). It simply is not possible with any realistic pseudogap model for there to be a pseudogap region of high curvature at 100 K as seen in the red data points in Fig.~\ref{chi}(a). The downward curvature would necessarily begin above 180 K. Such a narrow region of downturn can only be found at low temperature when $E^*$ is small, as seen in the black $p=0.182$ curve in Fig.~\ref{chi}(b) where $E^*/k_B$ = 51 K, or in the red dashed curve in Fig.~\ref{chi}(a) where $E^*/k_B$ = 8 K. Typically the region of maximum curvature lies near a temperature of about $E^*/k_B$ and the downward curvature extends up to $2E^*/k_B$. Thus the range is narrow only when $E^*$ is small. As noted, the identification of a $T^*$ simply from {\it any} downward curvature above $T_{\textrm{c}}$ is easily confused with superconducting fluctuations.

We make two further comments in relation to two prominent literature reports. Firstly, the Raman group of Sacuto {\it et al.} \cite{Sacuto2} very recently concluded that the pseudogap $T^*(p)$ line collapses vertically in the superconducting state but not until the strongly overdoped range $p = 0.224 \pm 0.002$ - just at the proposed location of the van Hove point. However, it is clear from the thermodynamic and superfluid density \cite{Anukool} measurements described above that the strongly entropy-depleting pseudogap is completely absent beyond $p^* = 0.19$ and these authors must be observing a spectral feature in the Raman response other than the pseudogap, with this feature disappearing at the Lifshitz transition. The discussion in the present paper strongly suggests that the Raman response for $p>0.19$ is influenced either by this proposed Lifshitz transition or by a partial gap at $E_F$ arising from superconducting fluctuations that is also seen in intrinsic tunnelling data \cite{Watanabe2,Benseman}.

Secondly, we have mentioned the back-bending pseudogap phase curve of He {\it et al.} \cite{He2}. These authors report a depletion of the integrated antinodal spectral weight just above $T_{\textrm{c}}$ that extends out to $p \approx 0.22$ and attributed by them to the pseudogap. At first sight this is consistent with the Sacuto Raman data \cite{Sacuto2}. However, it is clear from all the above that this is likely to be associated with loss of spectral weight arising from superconducting fluctuations just above $T_{\textrm{c}}$ \cite{Larkin,Tallon3}. Field-dependent measurements (or impurity substitution) would clarify this situation.

%For both Y$_{0.8}$Ca$_{0.2}$Ba$_2$Cu$_3$O$_{7-\delta}$ and Bi$_2$Sr$_2$CaCu$_2$O$_{8+\delta}$ the BCS condensation energy ratio, $U_0/(\gamma_n T_{c,mf}^2)$, remains constant across the overdoped regime, consistent with a single order parameter, then abruptly collapses as doping is reduced below $p = 0.19$ and the pseudogap opens \cite{Loram,Tallon2}. As noted this also coincides with a sudden Fermi surface transformation that occurs at precisely the same doping, as determined by quasiparticle scattering \cite{Davis}, and changes in Hall number \cite{Badoux,Storey2}. It is therefore incumbent on these authors to show how their Raman spectral feature is related to the well-established entropy-depleting pseudogap if they are to interpret the evolution of these features as pseudogap phenomenology.

\subsection*{4.2 The pseudogap `phase diagram'.} As a final topic we wish to discuss the individual data points in the pseudogap $T-p$ `phase diagram' of Vishik \cite{Vishik} and Legros \cite{Taillefer} as widely used by others. The $T^*$ data points from SIS tunneling in both figures are not temperatures but energy gaps divided by $k_B$. Further, these gaps in the overdoped region are superconducting gap magnitudes \cite{Tstar} and therefore unrelated to the pseudogap. We have already noted that the $T^* \approx 110$ K NMR data point in ref. \cite{Ishida} should instead be close to 0 K. Resistivity-derived $T^*$ values in Legros {\it et al.} are evaluated in the usual way from downturns in resistivity data taken from Oda {\it et al.} \cite{Oda}. However, closer scrutiny shows that the downturn for optimal doping should be lower (if treated consistently over all dopings) and the overdoped downturn is, again, attributable solely to superconducting fluctuations as described above.

The $c$-axis resistivity data points are due to Watanabe {\it et al.} \cite{Watanabe1} who find a semiconducting-like upturn on cooling below a certain $T^*(\rho_c)$ value. However, in an earlier publication these authors concluded in similar studies ``we find that the onset of the semiconducting $\rho_c(T)$ does not coincide with the opening of the spin gap seen in the $\rho_a(T)$" \cite{Watanabe2}. Note that within a tunnelling model for c-axis transport \cite{Watanabe2,Wittorff}, such an upturn is in fact anticipated with the onset of superconducting fluctuations due to the reduction in the electronic DOS. Moreover the highest doping value of $T^*$ was explicitly identified by these authors \cite{Watanabe1} as lying below the onset temperature for superconducting fluctuations, just as we have asserted.

Finally, the STS-derived $T^*$ data points in both Vishik and Legros are due to Gomes {\it et al.} \cite{Gomes} and these simply map out the onset of the depression in DOS caused by superconducting fluctuations, not by the closure of the pseudogap. These data points map nicely onto the superconducting `pairing temperature' inferred by Kondo {\it et al.} \cite{Kondo} and onto the doping-dependent mean-field transition temperature $T_{\textrm{c}}^{\textrm{mf}}(p)$ determined from an entropy conservation treatment of the fluctuation specific heat \cite{Tallon3}. In short, we believe these various reported $T^*(p)$ `phase diagrams' which merge with the $T_{\textrm{c}}(p)$ phase curve in the heavily overdoped region are incorrect and should be abandoned.

There are relatively few systematic studies which distinguish between the pseudogap and superconducting fluctuations, using for example an applied magnetic field, impurity substitution or by using a suitable fitting procedure for the overall $T$-dependence. Kokanovi\'{c} {\it et al.} \cite{Kokanovic} implement the latter, taking advantage of the broad temperature scale for the pseudogap compared with the relatively narrow domain of superconducting fluctuations. More precisely, plotting $\chi_c(T)-\chi_{ab}(T)$ versus $T$ (Fig. 3(a) of ref. \cite{Kokanovic}) eliminates an isotropic Curie term, $C/T$, and allows the diamagnetic fluctuation contribution, which is pronounced in $\chi_c(T)$, to be seen more clearly. Alloul {\it et al.} \cite{Alloul2} and Naqib {\it et al.} \cite{Naqib} apply a field to identify and suppress fluctuations. Both studies show that the $T^*(p)$ line cuts through the fluctuation pairing temperature above $T_{\textrm{c}}$, trending towards zero as $p \rightarrow 0.19$. However, the latter study, on epitaxial thin films of Y$_{0.8}$Ca$_{0.2}$Ba$_2$Cu$_3$O$_{7-\delta}$, was able to track $T^*$ below $T_{\textrm{c}}$ by combining progressive Zn substitution with their field studies. $T^*(p)$ was thereby found to cut the Zn-free $T_c(p)$ phase curve and continue undeflected towards zero as $p \rightarrow 0.19$. Even so we consider this as reflecting an underlying $p$-dependent energy scale which vanishes, rather than a closing temperature for the pseudogap.

A similar approach has been used by Usui {\it et al.} \cite{Usui} to study separate pseudogap and superconducting fluctuation effects in single-crystal Bi2212, using both in-plane and out-of-plane resistivity. They use the same method as we do to establish doping state. Using applied fields to identify the onset of a strong magneto-resistance they obtain a fluctuation range similar to what we have reported. They identify $T^*$ values from the usual downturn in resistivity up to a doping state of 0.16, but not at $p = 0.20$ where, already, the $T$-dependent resistivity is slightly superlinear. All this is consistent with the picture we have presented. However, they also extract $T^*$ values from the upturn in the $c$-axis resistivity, $\rho_c(T)$, which are $\approx 100$ K higher and which persist at least out to $p = 0.22$ and project to persist far beyond this. In our view, and as noted earlier, in the view of Watanabe {\it et al.} \cite{Watanabe2}, the upturn does not signal the opening of the pseudogap seen in the $a-b$-plane transport. Moreover, it is at odds with the later data reported by Watanabe \cite{Watanabe1}. And, in the recent ARPES study by the Z.-X. Shen group \cite{Chen} the pseudogap is only observed in the `incoherent strange metal state' which opens abruptly below $p \approx 0.19$ – thereby effectively abandoning their earlier reentrant scenario. This upturn in $\rho_c(T)$ which projects to persist outside of the superconducting dome cannot therefore be associated with the pseudogap.

A final, and central, question to consider is whether $p^*$ = 0.19 is a generic feature of the cuprates. The question is considered briefly in Appendix C and our conclusion is that, for {\it canonical cuprates}, it probably is general.

\section*{5. Conclusions}

In summary, in a search for either entrant or reentrant pseudogap behaviour we have examined the condensation free energy, $\Delta F(T,p)$, the magnetisation free energy, $\Delta F(H,T,p)$ and their entropy counterparts for Bi$_2$Sr$_2$CaCu$_2$O$_{8+\delta}$ in closely spaced doping states either side of critical doping, $p^* \approx 0.19$, where we find that the pseudogap closes. In every case we observe $\Delta F(T)$ to follow closely the near-weak-coupling $T$-dependence for a single unperturbed $d$-wave order parameter. There is no obvious enhancement in $\Delta F(T)$ on crossing a putative back-bending $T^*(p)$ line in the reentrant scenario, nor is there any obvious suppression in $\Delta F(T)$ on crossing a putative monotonically decreasing $T^*(p)$ line in the entrant scenario. One simply observes a strong reduction in the overall amplitude of the entire $\Delta F(T)$ curve once the pseudogap opens as $p$ falls below 0.19, with no change in its mean-field-like, single-order-parameter {\it shape}. Neither do we observe an associated inflexion anomaly in the entropy $\Delta S(T)$. Note that our main conclusions do not depend on the specific mean-field model used here for $\Delta F(T)$. For example, an alternative interpretation \cite{Benseman} involving pair-breaking below $T_{\textrm{c}}$ and Gaussian fluctuations above $T_{\textrm{c}}$ would also give smooth behavior in $\Delta F(T)$  i.e. no anomalies in the measured entropy, allowing us to rule out both entrant and reentrant behavior. In this alternative interpretation the values of $T_{\textrm{c}}^{\textrm{mf}}$ are much smaller, typically of order 1.1 and the remnant normal-state pseudogap proposed by others to extend to $p\approx 0.23$ \cite{Vishik,Sacuto2} is ascribed to Gaussian superconducting fluctuations at higher $T$ \cite{Loram5,Naqib2,Kokanovic2,Benseman} that cross over to critical fluctuations as $T_{\textrm{c}}$ is approached from above. Such a picture was proposed earlier \cite{Loram5} for the heat capacity of several cuprate families and may apply to the Bi2212 data discussed here. We conclude that $p^* \approx 0.19$ is the {\it temperature-independent} location where the pseudogap abruptly opens or closes, consistent with the observed fact that the pseudogap-induced lost entropy for $p<0.19$ is never recovered to well above room temperature. Paralleling this, the ``lost susceptibility" is never recovered, even to 500 K.  We reiterate that the pseudogap line often drawn on the phase diagram is actually the pseudogap energy scale (expressed as $E^*/k_B$) which falls with increasing doping and vanishes at $p^* = 0.19$. The pseudogap is still present and fully developed above this line to very high temperature. The thermodynamically-weak mean-field nematic transition observed near a sloping $T^*(p)$ line \cite{Sato} does not reflect the opening of the pseudogap but, rather, occurs within the preexisting pseudogap.

\section*{Appendix A: Thermodynamic functions}

\subsection*{A.1 Free energy} The differential specific heat technique allows the extraction of the electronic specific heat coefficient, $\gamma(T)$, from the much larger phonon term. From this we may obtain the electronic entropy by integration and the electronic free energy by integration again of the entropy.  We find it preferable to perform just a single integration and obtain the free energy using the following generic thermodynamic relation:
\begin{equation}
-\Delta F(T) =  \int_{\tau}^T \! T \, \Delta\gamma(T) \,\mathrm{d}T - T\,\int_{\tau}^T \!  \Delta\gamma(T) \,\mathrm{d}T ,
\label{free-energy}
\end{equation}
\noindent where the first term is the electronic internal energy, $-\Delta U(T)$, and the second term is the electronic entropy term, $T\Delta S(T)$. The term $\Delta \gamma(T)$ may refer to the difference between the superconducting and normal states i.e. $\Delta \gamma(T) = \gamma_{\textrm{n}} - \gamma_{\textrm{s}}$; $\Delta S(T) = S_{\textrm{n}} - S_{\textrm{s}}$ etc,  or it may refer to the difference in $\gamma$ between its value in field, $H$, and its value in zero field i.e. $\Delta \gamma(H,T) = \gamma(H,T) - \gamma(0,T)$; $\Delta S(H,T) = S(H,T) - S(0,T)$ etc. Ideally, the integrals in Eq.~\ref{free-energy} are from $\tau = T_{\textrm{c}}$ down to some value of $T < T_{\textrm{c}}$. However, the cuprates are distinguished by strong fluctuations over quite a broad range around $T_{\textrm{c}}$ and as a consequence the integrals must be performed from some temperature, $\tau$, well above $T_{\textrm{c}}$, sufficiently above the range of superconducting fluctuations, where $\Delta \gamma(\tau)$, $\Delta F(\tau)$ and $\Delta S(\tau)$ have fallen to zero.

\subsection*{A.2 Normal-state thermodynamic functions}
In order to determine the condensation terms $\gamma_{\textrm{n}} - \gamma_{\textrm{s}}$, $S_{\textrm{n}} - S_{\textrm{s}}$ and $F_{\textrm{n}} - F_{\textrm{s}}$ we need to evaluate the normal-state terms $\gamma_{\textrm{n}}(T)$ and $S_{\textrm{n}}(T)$. The former can be determined within fairly strict bounds by a simple construction of $\gamma_n(T)$ that satisfies entropy balance. However, we place the additional strict requirement that $\gamma_{\textrm{n}}(T)$ must be consistent with the DOS obtained from the electronic dispersion. We take the ARPES-derived dispersion of Kaminski {\it et al} \cite{Kaminski} and calculate the density of states $N(E)$ assuming the dispersion shifts rigidly (relative to the Fermi level) with doping. From the electronic density of states we calculate $S_{\textrm{n}}$ as described earlier \cite{Loram,Storey4,Tallon2} using the standard formula for weakly-interacting Fermions:
\begin{equation}
S_{\textrm{n}} = -2k_B \int_{-\infty}^\infty \! [f\ln(f) + (1-f)\ln(1-f)] \, N(E) \,\mathrm{d}E
\label{entropyeq}
\end{equation}
\noindent where $f(E)$ is the Fermi function.

\begin{table}
\centering
\begin{tabular}{|c|c|c|c|c|c|}
\hline
$p$  & $E_{\textrm{vHs}}-E_{\textrm{F}}$  & $E^*$ & $E^*$ & $\theta_{\textrm{FA}}$ & $T_{\textrm{c}}$ \\
(holes/Cu) & (meV) & (meV) & (meV) & ($\deg$) & (K) \\
 & & from DOS & from entropy & & \\ \hline
0.2093 & 8  & 0 & 0 & 0 & 71.2  \\  \hline
0.2030 & 14 & 0 & 0 & 0 & 75.6  \\  \hline
0.1944 & 18 & 0 & 0 & 0 & 81.4  \\  \hline
0.1879 & 23 & 0 & 1.3 & 0 & 83.9  \\  \hline
0.1821 & 35 & 4 & 4.4 & 10 & 85.8  \\  \hline
0.1758 & 55 & 12 & 8.8 & 22 & 87.4  \\  \hline
0.1694 & 65 & 18 & 12.3 & 25 & 88.6  \\  \hline
0.1616 & 76 & 22 & 15.8 & 28 & 89.3 \\  \hline
0.1527 & 84 & 24 & 18.5 & 30 & 89.1 \\  \hline
0.1382 & 91 & 30 & 23.8 & 34 & 85.6 \\  \hline
0.1288 & 107 & 36 & 27.8 & 35 & 81.9 \\  \hline

\end{tabular}
\caption{Fit parameters \cite{Storey4} to $S_{\textrm{n}}(T)/T$ using a rigid ARPES-derived dispersion \cite{Kaminski} showing doping state, $p$, distance from the van Hove singularity, $E_{\textrm{vHs}}-E_{\textrm{F}}$, pseudogap energy $E^*$ obtained from ARPES-derived DOS, $E^*$ obtained from entropy \cite{Loram}, Fermi arc angle, $\theta_{\textrm{FA}}$, which is a measure of the length of the Fermi arc, and $T_{\textrm{c}}$.  }
\label{Storeyparam}
\end{table}

The pseudogap is treated as before, both nodal \cite{Storey3} and with Fermi arcs \cite{Storey4}, and $S_{\textrm{n}}(T)$ is fitted to the normal-state experimental data, the only fitting parameters being the magnitude of the pseudogap, $E^*(p)$, the distance away from the van Hove singularity, $E_{\textrm{vHs}} - E_{\textrm{F}}$ and the Fermi arc angle $\theta_{\textrm{FA}}$ (when used). In the latter case the fits, carried out by JGS, are shown by the solid curves in Fig.~\ref{chi}(b) as reported earlier \cite{Storey4} and the fit parameters are listed in Table I. Also listed are the $E^*$ values determined previously from the downward (parallel) shift of the entropy $S(T)$ (for $T \gg T_{\textrm{c}}$) assuming the same triangular gap in the DOS mentioned in Section 4.1. The two value sets track each other very nicely despite the difference in the way in which the {\bf k}-dependence of the gap is described.

\begin{figure}
\centering
\includegraphics[width=75mm]{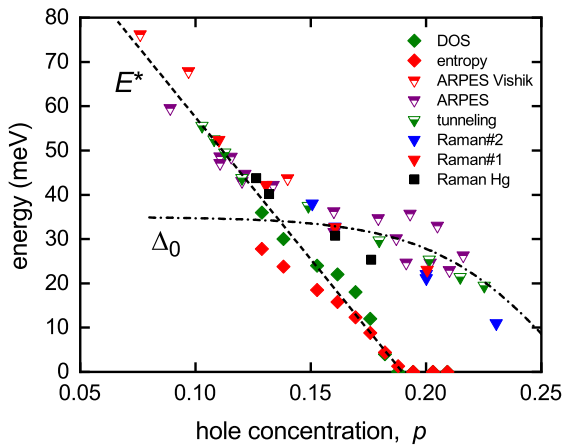}
\caption{\small
(Color online) The doping dependence of the antinodal gap as reported by Le Tacon {\it et al.} compared with our pseudogap fit values for $E^*$ from Table I. We have added the gap values reported by Vishik {\it et al.} \cite{Vishik}. The data project to $E^*(p=0) = 122$ meV, comparable to the nearest-neighbor exchange energy, $J$.
}
\label{Egplot}
\end{figure}

To compare with other reported measures of the pseudogap energy we reproduce in Fig.~\ref{Egplot} the data points reported in Fig. 2 from the Raman study of Le Tacon {\it et al.} \cite{LeTacon}. We use the same symbols and color schemes, but we omit the $B_{2g}$ data points as these relate to the superconducting gap only. To these we add the three antinodal $E^*$ gaps reported by Vishik {\it et al.} \cite{Vishik}. The data is low-temperature, well below $T_c$, and therefore the antinodal gap also includes the superconducting gap when it is comparable to, or greater than, $E^*$. To assist differentiation the dashed line shows the evolution of the pseudogap and the dash/dot curve that of $\Delta_0$. Our fit values are very consistent with this evolution. The pseudogap line projects to $E^*(p=0) = 122$ meV, consistent with the presumed underlying energy scale of $J$, the nearest-neighbour exchange interaction. We emphasise that measurements made above $T_c$ (and its fluctuation range) reveal the pseudogap only and the data points retreat to the dashed line.

The resulting data fits give an excellent description of the experimental entropy above $T_{\textrm{c}}$ and the values of $\gamma_{\textrm{n}}(T)$, obtained by differentiation, satisfy entropy conservation below $T_{\textrm{c}}$. (We note however that the precise location of the van Hove point differs from one cuprate to the next and even for Bi2212 its exact location does not affect our conclusions).

\subsection*{A.3 Superfluid density}
The field-dependent change of free energy $\Delta F(H,T)$ has been analysed by JLT and JWL in terms of the London model for the field dependent magnetisation \cite{Fetter}:
\begin{equation}
\mu_0 M(H,T) = \frac{\phi_0}{8\pi\lambda^2} \ln \left(2e^{2-2\gamma_E} \,\frac{H-H_{c1}}{H_{c2}}\right) ,
\label{magnetisation}
\end{equation}
\noindent where $\gamma_E$ is Euler's constant (=0.5772). By integration w.r.t. $H$ this yields:
\begin{equation}
\Delta F(H,T) = -\frac{\phi_0 (H-H_{c1})}{8\pi\lambda^2(T)} \ln \left(2e^{1-2\gamma_E} \,\frac{H-H_{c1}}{H_{c2}}\right) .
\label{HaoClem}
\end{equation}

Thus, by assuming $H \gg H_{c1}$ and plotting $\Delta F(H,T)/H$ versus $\ln(H)$ one expects linear behavior with slope proportional to the superfluid density, $\rho_s = \lambda^{-2}$, and $x$-axis intercept giving $\ln(H_{c2})$.

\subsection*{A.4 Spin susceptibility}
Eq.~\ref{entropyeq} may be regarded as an integral of the density of states weighted by a `Fermi window' function. The spin susceptibility is a similar thermodynamic function given by:
\begin{equation}
\chi_s = -2\mu_B^2 \int_{-\infty}^\infty \! \frac{\partial f(E)}{\partial E} \, N(E) \,\mathrm{d}E ,
\label{spineq}
\end{equation}
\noindent where the Fermi window is the function $\partial f/\partial E$. These two Fermi windows are essentially identical if the $\chi_s$ window is stretched in temperature by a factor 1.19 \cite{Loram6}. Thus, if the electronic system is weakly interacting, $S/T$ and $\chi_s$ will be related. In fact it was found for a number of cuprates that the normal-state $S/T$ is quantitatively very close to $a_W \chi_s$ over a very wide range of doping and temperature \cite{Loram1} consistent with the elementary excitations being weakly-interacting Fermions. Note that this is also true for many strongly correlated heavy-Fermion systems \cite{Jaklic}. Here $a_W$ is the Wilson ratio for weakly-interacting Fermions, $a_W = (\pi k_B/\mu_B)^2/(3\mu_0)$.

This being the case, we may apply tests to the spin susceptibility similar to those that we apply to the entropy, and $\chi_{\textrm{s}}$ may be expected to show superconducting fluctuations over a similar temperature range. We exploit this fact in Fig.~\ref{chi}.

\section*{Appendix B: Pseudogap closure at {\it \lowercase{p}}~$\approx~0.19$}
We have claimed since 1994 that the pseudogap closes abruptly at $p = 0.19 \pm 0.05$ \cite{Tallon1,Loram1,Loram3} for both Bi2212 and (Y,Ca)Ba$_2$Cu$_3$O$_{7-\delta}$. The independent evidence for this has grown substantially since then and we summarise it here, both for convenience and to show the compelling nature of this extensive evidence. As doping is reduced below this value we see an abrupt collapse in:

\noindent (i) the condensation energy \cite{Loram};

\noindent (ii) the jump $\Delta \gamma_c$ in specific heat coefficient at $T_{\textrm{c}}$ \cite{Loram} - see Fig.~\ref{gamma_jump};

\noindent (iii) the mean-field jump $\Delta \gamma^{\textrm mf}$ at $T_c^{\textrm mf}$ \cite{Tallon3} - see Fig.~\ref{gamma_jump};

\noindent (iv) the differential field-induced jump, $\Delta \gamma(H)$, at $T_{\textrm{c}}$ \cite{Loram} - see Fig.~\ref{gamma_jump} and Fig.~\ref{chi}(c);

\noindent (v) the ground-state superfluid density \cite{Anukool,Bernhard,Tallon6};

\noindent (vi) the entropy at $T_{\textrm{c}}$ \cite{Tallon6};

\noindent (vii) the Knight shift at $T_{\textrm{c}}$ \cite{Alloul,Storey3};

\noindent (viii) the critical impurity concentration to suppress $T_{\textrm{c}}$ to zero \cite{Tallon6};

\noindent (ix) the critical fields, $H_{c1}$, $H_c$ and $H_{c2}$ \cite{Grissonnanche,Tallon4}; and

\noindent (x) the self-field critical current, $J_c^{\textrm{sf}}$ \cite{Talantsev1,Kanigel,Tallon4}

\noindent - all signifying an abrupt crossover from strong superconductivity to weak superconductivity associated with removal of states or spectral weight available for superconductivity (note - this is not to be confused with ``weak-coupling" superconductivity).

\noindent (xi) This crossover is not merely of theoretical interest - these dramatic changes all combine to impact on the fine tuning of conductors for practical and commercial applications \cite{Talantsev1,Tallon4,Talantsev2}.

Below $p \approx 0.19$ there abruptly occurs:

\noindent (xii) a permanent (i.e. to very high temperature) loss of electronic entropy \cite{Loram};

\noindent (xiii) a permanent loss of spin susceptibility \cite{Naqib2,Loram};

\noindent (xiv) a large loss of spectral weight in the c-axis infrared conductivity extending to the highest temperatures investigated (300 K) \cite{Bernhard1};

\noindent (xv) an abrupt loss of ground-state ($\pi$,0) ARPES quasiparticle peak intensity that correlates with the loss of superfluid density and condensation energy \cite{Feng};

\noindent (xvi) a sudden change of the Fermi surface from large, with area $1+p$, to Fermi arcs or small hole pockets on the zone diagonal with area $p$ as seen in STS quasiparticle scattering \cite{Davis,Hudson} and ARPES \cite{Johnson,Damascelli};

\noindent (xvii) a normal-state crossover in Hall number $n_H$ from $1+p$ to $p$ \cite{Badoux,Storey1};

\noindent (xviii) a change in the sign of the Coulomb condensation energy occurring precisely at $p=0.19$ \cite{vanderMarel1,vanderMarel2};

\noindent (xix) a crossover from coherent to incoherent antinodal quasiparticles occurring precisely at $p=0.19$ independent of temperature and coinciding with the opening of the EDC pseudogap \cite{TallonMOS,Chen}.

\noindent (xx) a crossover from a near weak-coupling superconductor with $2\Delta_0/(k_BT_{\textrm{c}})$ = 4.5 across the overdoped region to a rapidly rising ratio in the underdoped region \cite{Chen}. Already by optimal doping, $p \approx 0.16$, the BCS ratio has doubled \cite{Chen}.

\noindent (xxi) The temperature-dependence of various normal-state properties at different doping states may be scaled onto a single curve using a doping-dependent scaling parameter $E^*(p)/k_B$ with the energy, $E^*(p)$ falling to zero at the critical doping $p \approx 0.19$. These include the in-plane thermoelectric power and the $c$-axis electrical resistivity \cite{Cooper2}; and

\noindent (xxii) the doping dependence for YBa$_2$Cu$_3$O$_{7-\delta}$ of the integrated weight of antiferromagnetic (AF) spin fluctuations $\int_{0}^{50 meV} \chi^{\prime\prime}({\bf q_{\textrm{AF}}},\omega) \,\mathrm{d}\omega$, as measured by inelastic neutron scattering, falls progressively with increasing doping to zero at $p = 0.19$ \cite{Tstar,Storey4}. This strongly suggests that the pseudogap is associated with short-range AF correlations.

At $p = 0.19$ various authors report:

\noindent (xxiii) $T$-linear normal state in-plane DC resistivity with sub-linear below and superlinear above \cite{Eckstein,TallonMOS}; and

\noindent (xxiv) very recently, Sterpetti {\it et al.} report a quantum-critical-like cusp-shaped crossover to a pseudogapped “strange metal phase” in single-unit-cell Bi2212. They conclude “we found it to be centered at about 0.19 holes/Cu” \cite{Sterpetti}.

\section*{Appendix C: Critical doping in other cuprates}
We have largely confined our discussion to Bi2212 because this was the subject material in the four key references \cite{Vishik,He2,Sacuto1,Taillefer} that we consider. Substantial evidence has been gathered in Appendix B for the rather abrupt closing of the pseudogap at $p^* = 0.19 \pm 0.005$ in both Bi2212 and Y123. Is this perhaps more general for all cuprates?

Turning then to other materials, determining the doping state of Bi$_2$Sr$_2$CuO$_6$ (Bi2201) has always been challenging. It does not follow the common relationship between doping and room-temperature thermopower \cite{Obertelli}. Thus see e.g. Ando {\it et al.} \cite{Ando}. The field has tended to use the method of Ando to specify doping for this system. However, it is noteworthy that the same crossover from a large Fermi surface to Fermi arcs is observed in STS quasiparticle scattering in both Bi2212 and Bi2201 (\cite{Davis} and \cite{Hudson}, respectively). For Bi2212 the doping of this crossover is determined by the Luttinger count to be $p \approx 0.19$. For Bi2201 the crossover occurs at $p \approx 0.15$ as determined by the Ando method but at $p \approx 0.19$ by the Luttinger count (Fig. 2(J) in \cite{Hudson}). If this is indeed the opening of the pseudogap then this specific location would indeed seem to be more general.

We are not aware of many relevant studies in the case of HgBa$_2$CuO$_{4+\delta}$ (Hg1201), however, we note that the broad features of the dome-shaped cuprate $T_{\textrm{c}}(p)$ phase diagram are dominated by the pseudogap. The superconducting energy gap amplitude $\Delta_0(p)$ typically falls along with $T_{\textrm{c}}(p)$ on the overdoped side but remains largely constant on the underdoped side \cite{Loram,Bernhard1,Pimenov,Vishik,He2}. As noted, the ratio 2$\Delta_0/(k_{\textrm{B}T_{\textrm{c}}})$ for Bi2212 remains more or less constant on the overdoped side with the value of $\approx4.5$ \cite{He2} while the ratio 2$\Delta_0/(k_{\textrm{B}}T_{\textrm{c}}^{\textrm{mf}})$ for both Bi2212 and (Y,Ca)123 is close to the weak-coupling value of 4.3 across the overdoped region \cite{Tallon3}. It is thus the opening of the pseudogap in the lightly overdoped region and its growth with underdoping which accounts for the underdoped fall in $T_{\textrm{c}}$ while $\Delta_0(p)$ remains fixed - see Fig. 4 of ref \cite{Tallon6}. Hence the well-known dramatic rise in the ratio 2$\Delta_0/(k_{\textrm{B}}T_{\textrm{c}})$ with underdoping \cite{He2}. Now (Y,Ca)123 and La214 have very similar $T_{\textrm{c}}(p)$ phase curves as a function of doping, differing only by an almost constant amplitude - see Fig. 2 of ref. \cite{Tallon9}. This implies the onset of the pseudogap close to $p \approx 0.19$ for La214 as in Y123. And significantly, Hg1201 also has a $T_{\textrm{c}}(p)$ phase curve almost identical to that of (Y,Ca)123 - see Fig. 5 of ref. \cite{Yamamoto} and c.f. Fig. 2 of ref. \cite{Tallon9}. This then suggests that the pseudogap opening for Hg1201 also lies near $p \approx 0.19$. Consistent with this Yamamoto {\it et al.} \cite{Yamamoto} present scaling plots of the resistivity and thermoelectric power which reveal a $T^*$ line that descends with doping very similar to that for (Y,Ca)123, projecting to a $p^*$ value close to 0.19, especially if the optimal-doped data is ignored where a $T^*$ value is difficult to ascertain as it approaches the value of $T_{\textrm{c}}$. All in all, the data is not inconsistent with $p^* \approx 0.19$ for Hg1201.

It follows from the above that the same can be said for La214. Significantly also, the $c$-axis infrared conductivity crosses over from insulating to metallic with a strong low-frequency Drude peak between doping states of 0.18 and 0.20 \cite{Henn}. Moreover, the jump height $\Delta \gamma(T_{\textrm{c}})$ begins to collapse rapidly below $p \approx 0.18$ \cite{Loram3}, as it does also for Tl$_{0.5}$Pb$_{0.5}$Sr$_2$Ca$_{1-x}$Y$_x$Cu$_2$O$_7$ \cite{Loram7}, in both cases indicating the abrupt opening of the pseudogap. Further, the phase behavior of Tl$_{0.5}$Pb$_{0.5}$Sr$_{2-x}$La$_x$CaCu$_2$O$_7$ is almost identical to that of Tl$_{0.5}$Pb$_{0.5}$Sr$_2$Ca$_{1-x}$Y$_x$Cu$_2$O$_7$ across the superconducting phase diagram \cite{Tallon10} suggesting that the pseudogap opens identically in this compound.

While more comprehensive studies would be desirable, this broad suite of cuprates all seem to exhibit an abrupt pseudogap opening somewhere near 0.19 holes/Cu. We may refer to these as {\it canonical} cuprates.  At the present time it seems unlikely that Tl$_2$Ba$_2$CuO$_{6+\delta}$ (Tl2201) would be different. For example, it's thermoelectric power \cite{Obertelli}, low-field Hall angle \cite{Mackenzie} and specific heat \cite{Loram8} show the usual canonical behavior. %Typically (perhaps always) it is overdoped. We attempted to underdope this system by reducing oxygen content and by La substitution on the Ba site but with no success. No pseudogap was observed in the specific heat \cite{Loram4}, even for the least overdoped composition. We are unaware of any definitive demonstration of the presence of the pseudogap in this system.
On the other hand, as noted in the introduction a crossover in Hall number from $p$ to $1+p$ was recently observed for Tl2201 \cite{Carrington} occurring as high as $p \approx 0.25$ and we have previously associated such a crossover with the closing of the pseudogap. That assignment is perhaps brought into question given the absence of a thermodynamic signature of the pseudogap in Tl2201 there. It has been shown theoretically \cite{Kivelsen} that a similar crossover in the Hall number is expected near a van Hove singularity if a Pomeranchuk-type nematic instability occurs, transforming a closed Fermi surface to open sheets. However, calculations (see Fig. S5 in ref. \cite{Carrington}) suggest that in Tl2201 the van Hove singularity lies at a much higher value of $p = 0.55$.

The other potential outlier is the cuprate system extensively studied by the group of Taillefer \cite{Taillefer2,Taillefer3}, namely La$_{1.6-x}$Nd$_{0.4}$Sr$_x$CuO$_4$. Drawing on transport and ARPES studies these authors identify the closing of the pseudogap at $p = 0.23 \pm 0.01$, well beyond the value for the other cuprates discussed above. It may well be that this system is non-canonical in this respect, however, we note that the ARPES investigation \cite{Matt} simply identified an antinodal gap "just above $T_{\textrm{c}}$" and associated this with the pseudogap. As discussed above it is now recognised that a pairing gap (or superconducting partial gap) persists well above $T_{\textrm{c}}$, in addition to the pseudogap over the lower doping range. The energy distribution curves reported for La$_{1.6-x}$Nd$_{0.4}$Sr$_x$CuO$_4$ just above $T_{\textrm{c}}$ are very similar to those reported for Bi2212 above $T_{\textrm{c}}$ \cite{Kondo} in terms of their shape and spectral-weight redistribution, but of lower resolution. A more extensive doping and temperature dependent study would be needed to truly ascertain that the observed partial gap is indeed the pseudogap.

The above is just a sketch of a crucially important topic which deserves much more detailed study. In principle we would expect that $p^* \approx N_0\times J$, where $J$ is the nearest-neighbor exchange interaction \cite{Loram} and $N_0$ is the density of states at the Fermi level. Inasmuch as both these quantities do not vary much from one cuprate to another we would not expect $p^*$ to vary much. This remains to be demonstrated as a canonical property of the cuprates and we encourage its investigation.

\begin{figure}
\centering
\includegraphics[width=70mm]{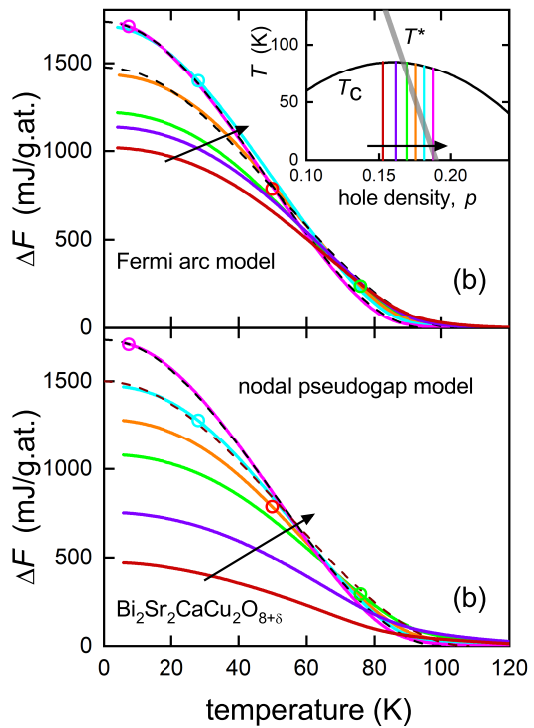}
\caption{\small
(Color online) Comparison of the effect on the deduced condensation energy, $\Delta F_{\textrm{ns}}$, of the two models used for the pseudogap: (a) Fermi arc model \cite{Storey4} and (b) nodal pseudogap model \cite{Storey3}. All color coding as in Fig. 2. Dashed curves are the near-weak-coupling $d$-wave $T$-dependence of $\Delta F_{\textrm{ns}}(T)$.
}
\label{FArc_vs_nodal}
\end{figure}

\section*{Appendix D: Comparison of Fermi arcs versus nodal pseudogap model}
We have used two models for the pseudogap in the past: a nodal pseudogap \cite{Storey3} on a background $S_{\textrm{n}}$ calculated from the rigid ARPES-derived dispersion of Kaminski {\it et al.} \cite{Kaminski} and a Fermi arcs model \cite{Storey4} based on the same background $S_{\textrm{n}}$. Beyond $p^*$ there is of course no impact so the calculated $\Delta F_{\textrm{ns}}$ values are as in Fig.~\ref{freeen}. It is useful to note the difference obtained between the two models for $p < p^*$ when the pseudogap is open. Clearly a nodal pseudogap strips away states available for superconductivity near the nodes so the condensation energy will be smaller than that arising in a Fermi arc picture. The two models are compared in Fig.~\ref{FArc_vs_nodal} and this confirms expectation. the differences are quite substantial. However, in either case a canonical near-weak-coupling $T$-dependence is still obtained (black dashed curves). The $\Delta F_{\textrm{ns}}$ curves for the nodal pseudogap were those reported previously \cite{Tallon2} for calculating the $T$-dependent superconducting gap amplitude. The Fermi arc derived $\Delta F_{\textrm{ns}}$ curves are new.

\bigskip
$^\dag$ jeff.tallon@vuw.ac.nz

$^\ddag$ deceased November 2017.

\end{document}